\documentclass{ws-ijmpcs}

\begin{document}

\markboth{Goldstein, Liuti, Gonzalez-Hernandez}
{Observations of Chiral Odd GPDs and their Implications}

%
\catchline{}{}{}{}{}
%

\title{OBSERVATIONS OF CHIRAL ODD GPD'S AND THEIR IMPLICATIONS 
}

\author{GARY R. GOLDSTEIN
}

\address{Department of Physics and Astronomy, Tufts University,\\
Medford, MA 02155,
USA
gary.goldstein@tufts.edu}

\author{SIMONETTA LIUTI}

\address{Department of Physics, University of Virginia, \\
Charlottesville, VA 22901, USA.\\
sl4y@virginia.edu}

\author{J. OSVALDO GONZALEZ-HERNANDEZ}

\address{Department of Physics, University of Virginia, \\
Charlottesville, VA 22901, USA.\\
jog4m@virginia.edu}

\maketitle

\begin{history}
\received{Day Month Year}
\revised{Day Month Year}
\end{history}

\begin{abstract}
Our original suggestion to investigate exclusive $\pi^0$ electroproduction  as a method
for extracting the tensor charge, transversity, and other quantities related to  chiral odd 
generalized parton distributions from cross section and asymmetry data is reviewed.   
We explain some of the details of the process: {\it i)} the connection between the helicity description and the cartesian basis; {\it ii)} the dependence on the momentum transfer squared, $Q^2$,  and {\it iii)} the angular momentum, parity, and charge conjugation
constraints ($J^{PC}$ quantum numbers). 

\end{abstract}

\ccode{PACS numbers: 11.25.Hf, 123.1K}
\vspace{0.2cm}

The theoretical framework for deeply virtual exclusive scattering processes using 
Generalized Parton Distributions (GPDs) has opened vast opportunities for understanding and interpreting hadron structure {\it including spin}  within QCD. Differently from both inclusive and semi-inclusive processes, GPDs can, in principle, provide essential information for determining the missing component of the nucleon  spin sum rule, which is identified with orbital angular momentum. 
However, a complete description of nucleon structure also requires the transversity (chiral odd/quark helicity-flip) GPDs, $H_T(x, \xi, t)$, $E_T (x, \xi, t)$, $\widetilde{H}_T (x, \xi, t)$, and $\widetilde{E}_T (x, \xi, t)$ \cite{Diehl_01}. 
Like the forward transversity parton distribution function, $h_1$, the transversity GPDs are expected to be more elusive quantities, not easily determined experimentally. 
It was nevertheless suggested in Ref.[\refcite{AGL}] that deeply virtual exclusive neutral pion electroproduction can provide a direct measure of chiral-odd GPDs so long as the helicity flip ($ \propto \gamma_5$) contribution to the quark-pion vertex is dominant. This coupling is subdominant, compared to the leading twist, chiral-even ($\propto \gamma_\mu\gamma_5$) contribution, but nevertheless the experimental data from both Jefferson Lab \cite{HallA,Kub} and HERA \cite{H1ZEUS} on meson electroproduction 
require such contributions.  Subsequently, other QCD practitioners \cite{GolKro} have pursued similar interpretations.

\begin{figure}
\includegraphics[width=7.cm]{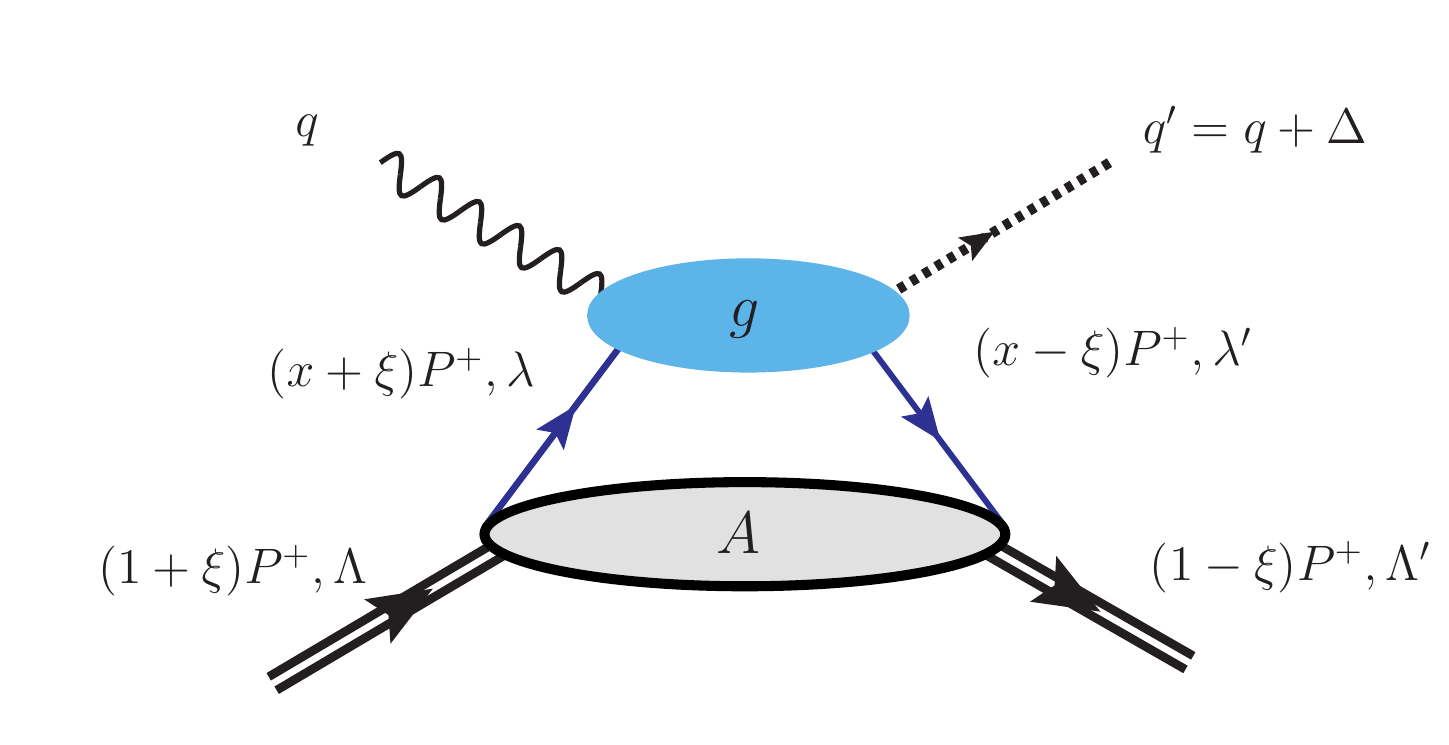}
\includegraphics[width=5.cm]{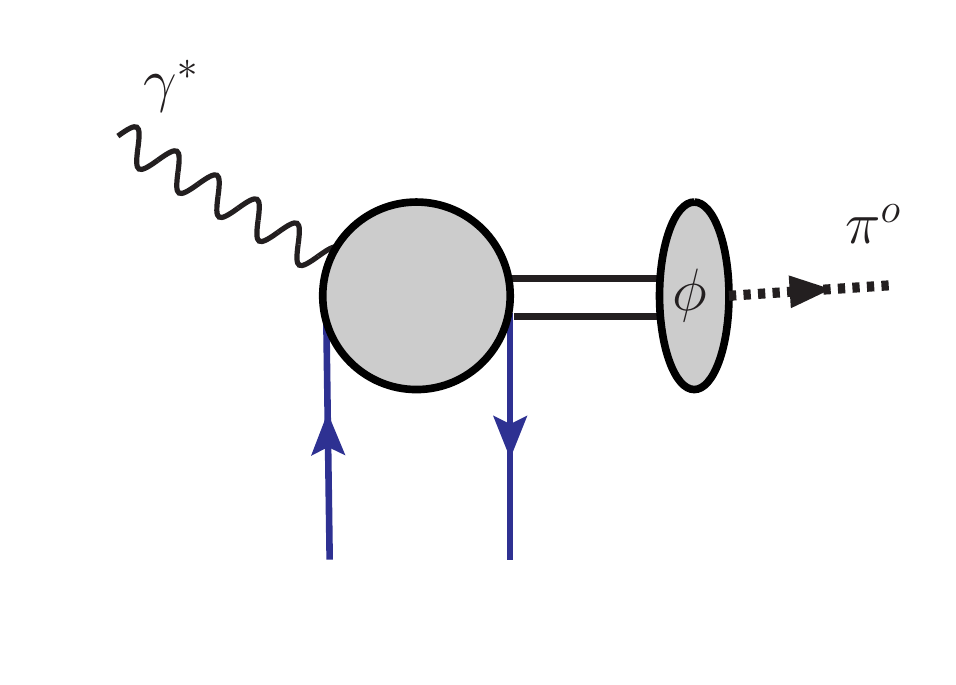}
\caption{Left: Leading order amplitude for DVMP, $\gamma^* + P \rightarrow M +P^\prime$. Notice that differently from Refs. \protect\refcite{GGL_even,GGL_odd}, we adopt the symmetric scheme of kinematics where $\overline{P}=(P+P')/2$ (see Ref.\protect\refcite{BelRad} for a review). Right: Hard scattering contribution to the DVMP, where $\phi$ is the outgoing meson distribution amplitude.}
\label{fig1b}
\end{figure}

Within a QCD framework at leading order the scattering amplitude factors into a nucleon-parton correlator described by GPDs, and a hard scattering part which includes the pion Distribution Amplitude (DA), Fig.~\ref{fig1b}. Keeping to a leading order description, the cross sections four-momentum squared dependences are $\sigma_L \approx {\cal O}(1/Q^6)$  and $\sigma_T \approx {\cal O}(1/Q^8)$, for the longitudinal and transverse virtual photon polarizations, respectively. It was expected that in the deep inelastic region $\sigma_L$ will be dominant.  Clearly, an explanation of data lie beyond the reach of the leading order collinear factorization~\cite{ColFraStr}. More interesting dynamics is involved. Consider the general form of $\pi$ coupling or distribution amplitude (DA) to quarks
\cite{Huang,BenFel}, 
\begin{eqnarray}
{\cal P} & = & K f_\pi \left\{ \gamma_5  \not\!{q}^\prime \phi_\pi(\tau) + \gamma_5 \mu_\pi  \phi_\pi^{(3)} (\tau) \right\}
\label{pi_coupling}
\end{eqnarray}
where $f_\pi$ is the pion coupling,  $\mu_\pi$ is a mass term that can {\it e.g.} be estimated from the gluon condensate, $\tau$ is the quark longitudinal momentum fraction, $\phi_\pi(\tau)$ and $\phi_\pi^{(3)} (\tau)$ are the naive twist-2 and twist-3 pion DAs. These determine chiral even and chiral odd processes.
In Ref.\refcite{AGL} it was seen that sufficiently large transverse cross sections can be produced in $\pi^0$ electroproduction  provided the chiral odd coupling is  adopted. 

The amplitude for Deeply Virtual Meson Production (DVMP) off a proton target is written  in a QCD factorized picture, using the helicity formalism, as the following convolution over the quark momentum components \cite{AGL,GGL_even}, 
\begin{eqnarray}
f_{\Lambda_\gamma,\Lambda;0,\Lambda^\prime}(\xi,t) & = & \sum_{\lambda,\lambda^\prime} 
\int dx d^2k_\perp \; g_{\Lambda_\gamma,\lambda;0,\lambda^\prime} (x, k_\perp, \xi, t)  \,
A_{\Lambda^\prime,\lambda^\prime;\Lambda,\lambda}(x, k_\perp, \xi,t)
\label{facto}
\end{eqnarray}
where the variables $x, \xi, t$ are $x=(k^+ + k^{\prime \, +})/(P^+ + P^{\prime \, +})$, $\xi=\Delta^+/(P^+ + P^{\prime \, +})$, $t=\Delta^2$; the particles momenta and helicities,  along with the hard scattering amplitude, $g_{\Lambda_\gamma,\lambda;0,\lambda^\prime}$, and the quark-proton scattering amplitude, $A_{\Lambda^\prime,\lambda^\prime;\Lambda,\lambda}$, are displayed in Fig.\ref{fig1b}. The $Q^2$ dependence is omitted for ease of presentation.

In our approach the chiral even contribution is sub-leading.
As explained in detail in Refs.\refcite{AGL,GGL_odd}, for the chiral odd contribution to $\pi^0$ electroproduction only $g_{1+,0-} \equiv g_T$, and $g_{0+,0-} \equiv g_L$, are different from zero. Furthermore $g_L$ is suppressed at ${\cal O}(k'_\perp/Q)$. Their expressions are
\begin{eqnarray}
\label{gT_odd}
g_T^{A,V} & = &  g_\pi^{odd}(Q) \left[\frac{1}{x-\xi + i \epsilon} -  \frac{1}{x + \xi - i \epsilon}  \right] = g_{\pi \, odd}^{A,V}(Q) \; C^- \\
\label{gL_odd}
g_L^{A,V} & = &   g_\pi^{odd}(Q)  \sqrt{\frac{t_o-t}{Q^2}} \left[  \frac{1}{x-\xi + i \epsilon} -  \frac{1}{x +\xi - i \epsilon} \right] =  
g_{\pi \, odd}^{A,V}(Q)  \sqrt{\frac{t_o-t}{Q^2}}   \; C^-,
\end{eqnarray}
where $g_{\pi \, odd}^{A,V}(Q)$ describe the pion vertex. The labels $A,V$ describe axial-vector and vector t-channel quantum numbers corresponding to exchanges as we will explain below. 
Notice that the coefficients $C^-$ from the quark propagator is even under crossing.
Using the allowed hard scattering amplitudes we obtain the following six independent amplitudes, of which four are for the transverse photon,
\begin{subequations}
\label{f_T}
\begin{eqnarray}
\label{f1}
f_1 = f_{1+,0+} &  = & g_{1+,0-} \otimes A_{+-,++}  \\
\label{f2}
f_2 = f_{1+,0-} &  = &  g_{1+,0-}  \otimes  A_{--,++}     \\
\label{f3}
f_3  = f_{1-,0+} & = &  g_{1+,0-}  \otimes  A_{+-,-+}    \\  
\label{f4}
f_4  = f_{1-,0-} & = &   g_{1+,0-}   \otimes  A_{--,-+},  
\end{eqnarray}
\label{chiral_odd_T}
\end{subequations}
and two for the longitudinal photon,
\begin{subequations}
\label{f_L}
\begin{eqnarray}
\label{f5}
f_5 = f_{0+,0-} &  = & g_{0+,0-}  \otimes  A_{--,++}    \\
\label{f6}
f_6 = f_{0+,0+} &  = & g_{0+,0-}  \otimes  A_{+-,++}, 
\label{chiral_odd_L}
\end{eqnarray}
\end{subequations}
The unpolarized target $\pi^0$ electroproduction cross section is given by,
\begin{eqnarray} 
\label{xs}
\frac{d^4\sigma}{d\Omega d\epsilon_2 d\phi dt} & = &\Gamma \left\{ \frac{d\sigma_T}{dt} + \epsilon_L \frac{d\sigma_L}{dt} + \epsilon \cos 2\phi \frac{d\sigma_{TT}}{dt} 
+ \sqrt{2\epsilon_L(1+\epsilon)} \cos \phi \frac{d\sigma_{LT}}{dt}  \right. \nonumber \\
& + & \left.
h  \, \sqrt{2\epsilon_L(1-\epsilon)} \, \frac{d\sigma_{L^\prime T}}{dt} \sin \phi \right\}, 
\end{eqnarray}
where, $\Gamma=(\alpha/2\pi^2) (k_e'/k_e)(k_\gamma/Q^2) /(1-\epsilon)$, with $k_e, k_e'$ being the initial and final electron energies, $k_\gamma$ the real photon equivalent energy in the lab frame, $h=\pm1$ the electron beam polarization, $\epsilon$ and $\epsilon_L=Q^2/\nu^2 \epsilon$, are the transverse and longitudinal polarization fractions, respectively \cite{DreTia,DonRas}, and the cross sections terms read,
\begin{subequations}
\label{xsecs}
\begin{eqnarray}
\frac{d\sigma_T}{dt} & = & \mathcal{N}  \, \left( \mid f_1 \mid^2 + \mid f_2 \mid^2 + \mid f_3 \mid^2 +
\mid f_4 \mid^2 \right)  
\\
\frac{d\sigma_L}{dt} & = & \mathcal{N} \,  \left( \mid f_5 \mid^2 + \mid f_6 \mid^2 \right),  \\
\label{dsigTT}
\frac{d\sigma_{TT}}{dt} & = & 2 \, \mathcal{N} \,    \Re e \left( f_1^*f_4 - f_2^* f_3 \right). \\
\frac{d\sigma_{LT}}{dt} & = & 2 \, \mathcal{N} \, 
\Re e \left[ f_5^* (f_2 + f_3) + f_6^* (f_1 - f_4) \right]. \\
\frac{d\sigma_{LT^\prime}}{dt} & = & 2 \, \mathcal{N}  \, 
\Im m  \left[ f_5^* (f_2 + f_3)  + f_6^* (f_1 - f_4) \right]
\end{eqnarray}
\end{subequations}
with $\mathcal{N}=(\pi/2)/[s (s-M^2)]$.  
In order to understand how the chiral odd GPDs enter the helicity structure we need to make a connection between the helicity amplitudes in Eqs.(\ref{f_T},\ref{f_L}), and the four Lorentz structures for the hadronic tensor first introduced in Ref.\refcite{Diehl_01}, 
\begin{eqnarray}
\label{CFF_odd}
&& \epsilon^\mu_T T_\mu^{\Lambda \Lambda^\prime}  =    e_q \int_{-1}^1  dx \:   
\frac{g_{T}}{2 \overline{P}^+} \; {\overline{U}(P',\Lambda')}\left[ i \sigma^{+i} H_T^q(x,\xi,t) +   
\frac{\gamma^+ \Delta^i - \Delta^+ \gamma^i}{2M} E_T^q(x,\xi,t) \right.  \nonumber \\
&& \left. \frac{\overline{P}^+ \Delta^i - \Delta^+ \overline{P}^i}{M^2}  \widetilde{H}_T^q(x,\xi,t) +
\frac{\gamma^+ \overline{P}^i - \overline{P}^+ \gamma^i}{2M} \widetilde{E}_T^q(x,\xi,t) \right] U(P,\Lambda),
\end{eqnarray}
where $i=1,2$, and $\epsilon^\mu_T$ is the transverse photon polarization vector. 

The GPD content of each amplitude therefore is,
\begin{subequations}
\label{f_amps_1}
\begin{eqnarray}
f_1 &= &  g_{\pi \, odd}^V(Q) \frac{\sqrt{t_0-t}}{2M(1+\xi)^2} \left[ 2\widetilde{\cal H}_T  + (1-\xi)  \left({\cal E}_T - \widetilde{\cal E}_T \right) \right] \\
f_2 & = &    (g_{\pi, \, odd}^V(Q) + g_{\pi, \, odd}^A(Q))  \,  \frac{\sqrt{1-\xi^2}}{(1+\xi)^2}  \left[ {\cal H}_ T + \frac{t_0-t}{4M^2} \widetilde{\cal H}_T  \right. \nonumber \\    
& & \left. + \frac{\xi^2}{1-\xi^2}  {\cal E}_T  + \frac{\xi}{1-\xi^2} \widetilde{\cal E}_T \right]  \\
\nonumber \\ 
f_3  & = &   (g_{\pi, \, odd}^V(Q) - g_{\pi, \, odd}^A(Q))  \,   \frac{\sqrt{1-\xi^2}}{(1+\xi)^2}  \,  \frac{t_0-t}{4M^2} \, \widetilde{\cal H}_T  
 \\
f_4 & = &  g_{\pi \, odd}^V(Q) \,  \frac{\sqrt{t_0-t}}{2M(1+\xi)^2}  \,  \left[  2\widetilde{\cal H}_ T + (1+\xi) \left( {\cal E}_T +
 \widetilde{\cal E}_T \right) \right]. 
\end{eqnarray} 
\end{subequations}
where we write,
\[ {\cal F}_T(\xi,t,Q^2) = \int_{-1}^1 dx \; C^- \, F_T(x,\xi,t,Q^2) \; \; \; \; \;  F_T \equiv {\cal H}_T, {\cal E}_T, \widetilde{\cal H}_ T, \widetilde{\cal E}_ T. \]  
These $Q^2$ dependent coefficients are derived according to the values of the $-t$-channel $J^{PC}$  quantum numbers for the process. They describe an angular momentum dependent formula that we discuss below. The amplitudes for longitudinal photon polarization, $f_5$ and $f_6$ are obtained similarly, by working out  Eqs.(\ref{f5},\ref{f6}) (details are given in Ref.\refcite{GGL_odd}).
The quark flavor content  of the quark-proton helicity amplitudes for $\pi^0$ electroproduction is, 
\begin{equation}
A_{\Lambda^\prime,\lambda^\prime;\Lambda,\lambda} = e_uA^u_{\Lambda^\prime,\lambda^\prime;\Lambda,\lambda}-e_dA^d_{\Lambda^\prime,\lambda^\prime;\Lambda,\lambda}. 
\end{equation}
 
Taking the dominant terms in $Q^2$ in Eqs.(\ref{f_amps_1}), and small $\xi$ and $t$  gives the dominant GPD contributions to Eqs.(\ref{xsecs}) 
\begin{eqnarray}
\frac{ d \sigma_T }{dt } & \approx & {\cal N}  \;  \left[
\mid {\cal H}_T \mid^2 + \,  \tau \left( \mid  \overline{\cal E}_T  \mid^2 + \mid  \widetilde{\cal E}_T  \mid^2 \right)  \right]   \\
\frac{ d \sigma_L }{dt } & \approx & {\cal N}  \; \frac{2M^2  \tau}{Q^2}  \mid {\cal H}_T \mid^2 \\
\frac{ d \sigma_{TT} }{dt } & \approx   &  {\cal N}     \;  \tau
\left[ \mid \overline{\cal E}_T \mid^2 -\mid  \widetilde{\cal E}_T \mid^2   + \;
\Re e  {\cal H}_T  \,  \frac{\Re e (\overline{\cal E}_T   - {\cal E}_T)}{2}  + \Im m  {\cal H}_T  \,  \frac{\Im m  (\overline{\cal E}_T   -  {\cal E}_T)}{2}     \right]  \nonumber \\
&& \\
\frac{ d \sigma_{LT} }{dt } & \approx &  {\cal N}    \; 2 \, \sqrt{ \frac{2M^2  \, \tau}{Q^2}}  \mid {\cal H}_T \mid^2 \\
\frac{ d \sigma_{L^\prime T} }{dt } & \approx & {\cal N} \; \tau \, \sqrt{\frac{2 M^2 \, \tau}{Q^2} } \;  \left[  \Re e  {\cal H}_T  \,   \frac{\Im m (\overline{\cal E}_T   -  {\cal E}_T)}{2}   -   \Im m  {\cal H}_T  \,   \frac{\Re e   (\overline{\cal E}_T   -  {\cal E}_T)}{2}  \right]
\end{eqnarray}
where $\tau= (t_o-t)/2M^2$, and we have redefined $\overline{\cal E}_T = 2 {\cal H}_T + {\cal E}_T$ according to Ref.[\refcite{Bur2}]. 

\begin{figure}
\centerline{
\includegraphics[width=6.5cm]{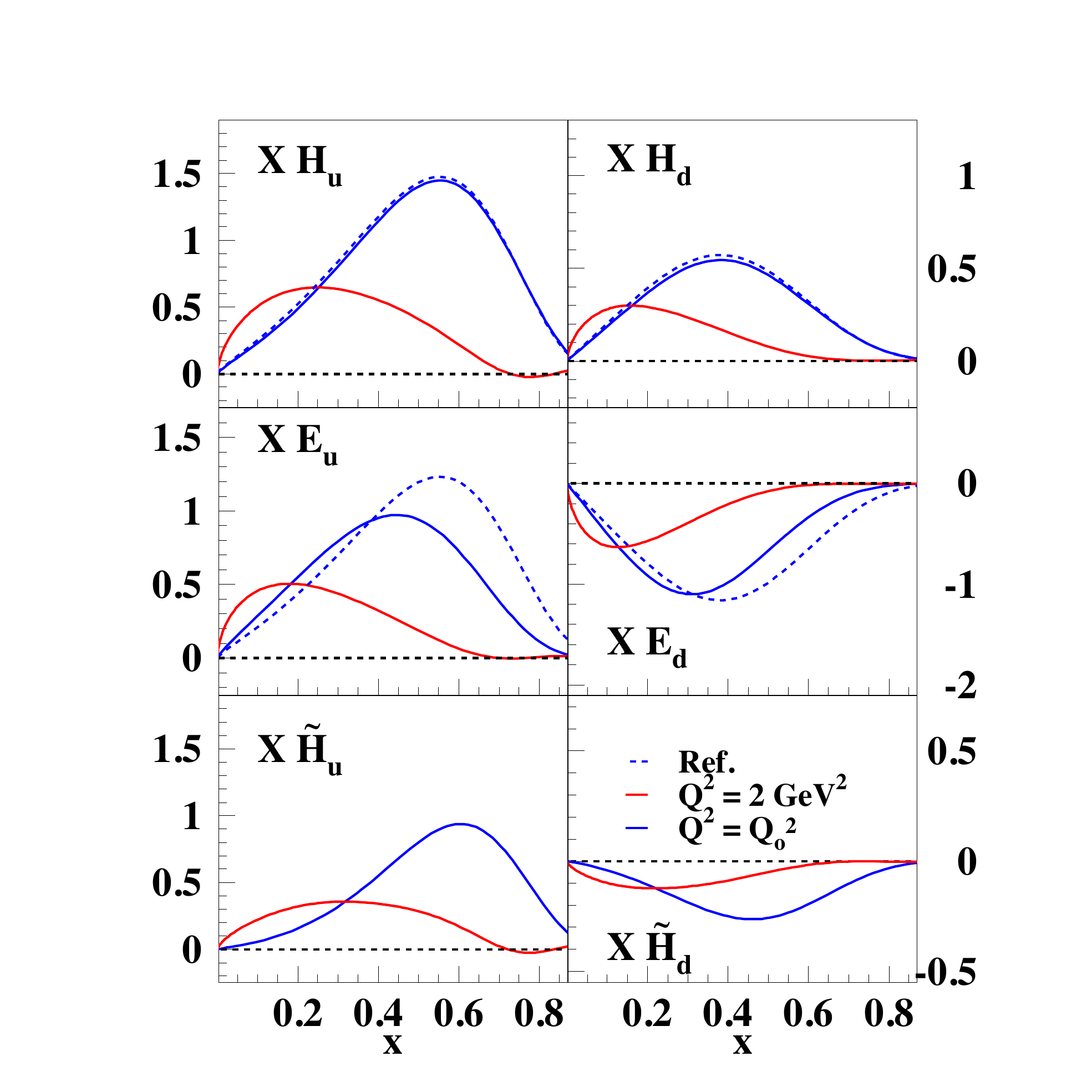}}
\caption{GPDs $H_q, E_q, {\widetilde E}_q$ for flavor q = u (left) and d (right) evaluated at an initial scale and $Q^2=2$ GeV$^2$, respectively. 
\protect\cite{GGL_even}.}
\label{chiral_even}
\end{figure}

Very little is known about the size and overall behavior of the chiral odd GPDs, besides that $H_T$ becomes the transversity structure function, $h_1$, in the forward limit,  $\overline{E}_T$'s first moment can be thought of as  the proton's ``transverse anomalous magnetic moment" \cite{Bur2},  and $\widetilde{E}_T$'s first moment is null \cite{Diehl_01}. 
To evaluate the chiral odd GPDs in Ref. \refcite{GGL_odd} we propose a method that, by using Parity transformations in a spectator picture, allows us to write them as linear combinations of the better determined chiral even GPDs as we determined in Ref.\refcite{GGL_even}. In the latter we introduced a ``flexible parameterization" for the GPDs that has both the diquark spectator structure and the Regge behavior for small $x$. See Fig.~\ref{chiral_even} for examples of the information obtained for the chiral even GPDs - as determined from our recursive fitting procedure. Our chiral even GPD parameterization also determines the electromagnetic form factors of the nucleons (Ref.~\refcite{FF}), in agreement with recent precise measurements~\cite{Cates}.

Based on our analysis we expect the following behaviors to approximately appear in the data:

\noindent {\it i)} The order of magnitude of the various terms approximately follows a sequence determined by the inverse powers of $Q$ and the powers of $\sqrt{t_o-t}$: $d \sigma_T/dt \geq d \sigma_{TT}/dt \geq d \sigma_{LT/L'T}/dt \geq d \sigma_L/dt; $

\noindent {\it ii)} $d \sigma_T/dt$ is dominated by ${\cal H}_T$ at small $t$, and governed by the interplay of  ${\cal H}_T$ and $\overline{\cal E}_T$ at larger $t$; 

\noindent {\it iii)} $d \sigma_L/dt$ and $d \sigma_{LT}/dt$ are directly sensitive to ${\cal H_T}$; 

\noindent {\it iv)} $d \sigma_{TT}/dt$ and $d \sigma_{L'T}/dt$ contain a mixture of GPDs. They will play an important role in 
singling out the less known terms, $\overline{E}_T$, $E_T$, and $\widetilde{E}_T$. 

\begin{figure}
\begin{center}
\includegraphics[width=7.5cm]{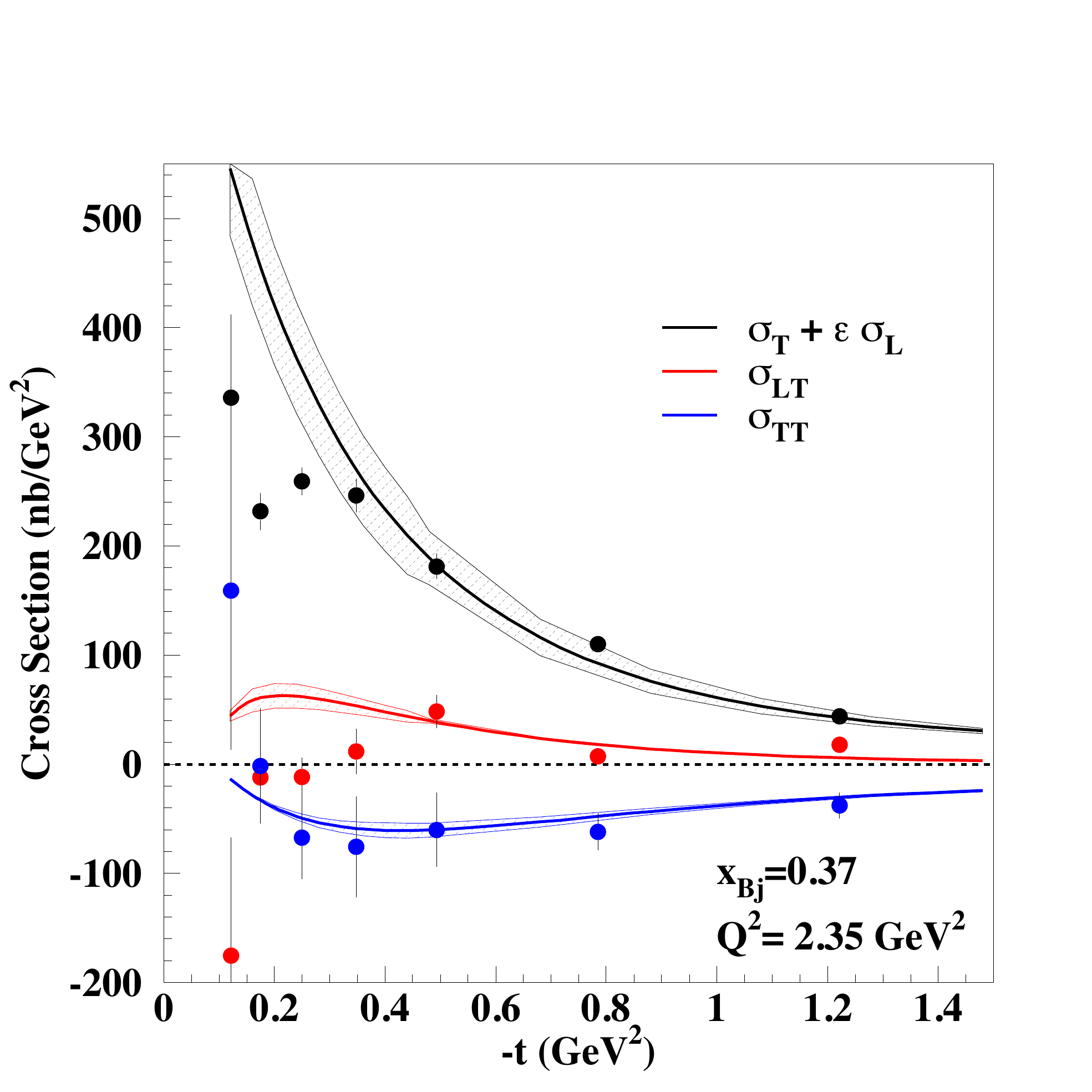}
\end{center}
\caption{(color online) $d \sigma_T/dt + \epsilon_L d \sigma_L/dt$, $d \sigma_{TT}/dt$, and $d \sigma_{LT}$ calculated using the physically motivated parametrization of Ref.\protect\refcite{GGL_odd}, plotted vs. $-t$ for $x_{Bj}=0.37$, and $Q^2=2.35$ GeV$^2$, along with data from Hall B \protect\cite{Kub}. The hatched areas represent the theoretical errors for the parametrization, which are mostly originating from the fit to the nucleon form factors \protect\cite{GGL_even,GGL_odd}. See Ref.\protect\refcite{Kub} for a full range of kinematic values.}
\label{fig2}
\end{figure}

\vspace{0.3cm}
\noindent The interplay of the various GPDs can already be seen by comparing to the Hall B data \cite{Kub} shown in  in Fig.\ref{fig2}. 
One can see, for instance, that the ordering predicted in {\it i)} is followed, and that  $d \sigma_T/dt$ exhibits a form factor-like fall  off of ${\cal H}_T$ with $-t$. 

As we explain in Ref.~\refcite{GGL_odd}, there exist two distinct series  of $J^{PC}$ configurations in the $t$-channel, namely the {\it natural parity} one ($1^{--}, 3^{--} ... $), labeled $V$, and 
the {\it unnatural parity} one ($1^{+-}, 3^{+-} ...$), labeled $A$. We hypothesize that  the two series will generate different contributions to the pion vertex. 
We consider separately the two contributions $\gamma^* (q \bar{q})_V \rightarrow \pi^0$ and 
$\gamma^* (q \bar{q})_A \rightarrow \pi^0$  to the process in Fig.~\ref{fig1b}.  
What makes the two contributions distinct is that, in the natural parity case (V), the orbital angular momentum, $L$, is the same for the initial and final states, or $\Delta L=0$,
while for unnatural parity (A), $\Delta L =1$. 
We modeled this difference  with the following expressions containing a modified kernel 
\begin{eqnarray}
g^V_{\Lambda_{\gamma^*},\lambda; 0, \lambda^\prime} & =  & \int dx_1 dy_1 \int  d^2 b  
\, \hat{\psi}_V(y_1,b) \, \hat{{\cal F}}_{\Lambda_{\gamma^*},\lambda; 0, \lambda^\prime}(Q^2,x_1,y_1,b) \alpha_S(\mu_R)
 \exp[-S]     \, \hat{\phi}_{\pi^o}(x_1,b)  \nonumber \\
 & = & g_{\pi \, odd}^{V}(Q) \; C^- \\
g^A_{\Lambda_{\gamma^*},\lambda; 0, \lambda^\prime}  & = &  \int dx_1 dy_1 \int d^2  b  
\, \hat{\psi}_A(y_1,b) \, \hat{{\cal F}}_{\Lambda_{\gamma^*},\lambda; 0, \lambda^\prime}(Q^2,x_1,y_1,b) \alpha_S(\mu_R)
\exp[-S]  \, \hat{\phi}_{\pi^o}(x_1,b) \nonumber \\
& = & g_{\pi \, odd}^{A}(Q) \; C^-
\end{eqnarray}
where, 
\begin{equation}
\hat{\psi}_{A}(y_1,b) = \int d^2 k_T J_1(y_1 b) \psi_V(y_1,k_T) 
\end{equation}
Notice that we now have an additional function, $\hat{\psi}_{V(A)}(y_1,b)$ that takes into account the effect of 
different $L$ states. The higher order Bessel function describes the situation where $L$ is always larger in the initial state. 
In impact parameter space this corresponds to configurations of larger radius. 
The matching of the $V$ and $A$ contributions to the helicity amplitudes is as follows: $f_1, f_4 \propto g^V$, $f_2 \propto g^V+g^A$, 
$f_3 \propto g^V-g^A$, thus explaining the $Q^2$-dependent factors in Eqs.(\ref{f_amps_1}).

\vspace{0.2cm}
\noindent {\bf Conclusions}

In summary, we introduced a mechanism for the $Q^2$ dependence of the process $\gamma^* q \bar{q} \rightarrow \pi^o$, that distinguishes among natural and unnatural parity configurations. We tested the impact of this mechanism using  the modified perturbative approach as a guide. Other schemes could be explored  \cite{Rad}.

Exclusive pseudoscalar meson electroproduction is directly sensitive to leading twist chiral-odd GPDs which can at present be extracted from available data in the multi-GeV region. We examined and justified this proposition further by carrying out a careful analysis of the chiral odd sector.
After going through a step by step derivation of  the connection of the helicity amplitudes formalism with the cartesian basis, we showed in Ref.~\refcite{GGL_odd}  how the dominance of the chiral-odd process follows unequivocally owing to the values allowed for the $t$-channel spin, parity, charge conjugation and from the GPDs crossing symmetry properties.
This observation has important consequences for the the $Q^2$-dependence of the process. Our $J^{PC}$ analysis supports 
the separation between the $Q^2$ dependence of the photo-induced transition functions 
for the even and odd parity combinations into $\pi^0$, thus
reinforcing the idea that spin related observables exhibit a non-trivial asymptotic behavior. The various beam and target asymmetries that further select independent combinations of the GPDs will be presented in the near future~\cite{hybrid_odd}.
A separation of the various chiral-odd GPDs contributions can be carried out, provided an approach that allows one to fix their parameters and normalizations, appropriately, is adopted like the one we presented here.

\vspace{0.5cm}
We thank the organizers of QCD Evolution Workshop 3 and P. Kroll, V. Kubarovsky, M. Murray, A. Radyushkin, P. Stoler, C. Weiss, for many interesting discussions and constructive remarks.  This work was partially supported by the U.S. Department
of Energy grants DE-FG02-01ER4120 (J.O.G.H., S.L.),  DE-FG02-92ER40702  (G.R.G.).


\end{document}